\documentclass[3p,times,procedia]{elsarticle}
\usepackage{nupha_ecrc}

\volume{00}

\firstpage{1}

\journalname{Nuclear Physics A}

\runauth{}

\jid{nupha}

\jnltitlelogo{Nuclear Physics A}

\usepackage{amssymb}

\usepackage[figuresright]{rotating}

\newcommand{\pp}{$\rm{p}+\rm{p}$}

\newcommand{\AuAu}{$\rm{Au}+\rm{Au}$}

\newcommand{\sqrtsNN}{\ensuremath{\sqrt{s_{\rm{NN}}}}}
\newcommand{\gev}{GeV/$c$}
\newcommand{\gevtwo}{GeV/$c^{2}$}
\newcommand{\jpsi}{\ensuremath{J/\psi}}
\newcommand{\ups}{\ensuremath{\Upsilon}}
\newcommand{\pT}{\ensuremath{p_{\rm{T}}}}
\newcommand{\RAA}{\ensuremath{R_{\rm{AA}}}}
\newcommand{\vtwo}{\ensuremath{v_{2}}}
\newcommand{\ncoll}{\ensuremath{N_{\rm{coll}}}}
\newcommand{\npart}{\ensuremath{N_{\rm{part}}}}

\begin{document}

\begin{frontmatter}



\dochead{}

\title{\jpsi\ and \ups\ measurements via the di-muon channel in \AuAu\ collisions at \sqrtsNN\ = 200 GeV with the STAR experiment}


\author{Rongrong Ma, on behalf of the STAR Collaboration}

\address{Brookhaven National Laboratory, Upton, NY 11973, USA}

\begin{abstract}
Measurements of quarkonium production in heavy-ion collisions have played an essential role in understanding the properties of the Quark Gluon Plasma created in such collisions. In early 2014, the Muon Telescope Detector, designed to trigger on and identify muons based on its precise timing information, was fully installed in STAR. It opens the door to measure quarkonia via the di-muon channel for the first time at the STAR experiment, with the potential to separate different \ups\ states. In this talk, we present the measurements of \jpsi\ suppression and elliptic flow at mid-rapidity in Au+Au collisions at \sqrtsNN\ = 200 GeV down to low transverse momentum (\pT). The suppression is found to decrease with increasing \pT\ while the elliptic flow is consistent with 0 for \pT\ above 2 \gev. Furthermore, the measurement of different \ups\ states is explored within the precision of the available statistics.
\end{abstract}


\end{frontmatter}


\section{Introduction}
\label{sect::intro}
Colliding heavy ions at relativistic energies has been used to create in laboratory a new state of matter, usually referred to as the ``Quark Gluon Plasma (QGP)", where the decomfined quarks and gluons are the relevant degrees of freedom. Exploring the properties of such matter under extreme conditions provides new tests to Quantum Chromodynamics (QCD). Among various probes, suppression of quarkonia, especially \jpsi, due to the dissociation of the bound states by the color-screening of the surrounding partons in the medium was proposed as a direct proof of the QGP existence \cite{Matsui:1986dk}. However, recombination of un-correlated charm and anti-charm quarks in the medium complicates the interpretation of \jpsi\ suppression. Measurement of second-order harmonic coefficient (``elliptic flow" or \vtwo) for \jpsi\ would help disentangle different production mechanisms since the regenerated \jpsi\ is expected to inherit the flow of the charm quarks while the primordial \jpsi\ should have almost zero \vtwo\ \cite{Adamczyk:2012pw}. On the other hand, contribution to the production of \ups\ states from regeneration is negligible at RHIC energies, making them ideal probes to study the color-screening feature.

A new Muon Telescope Detector (MTD) \cite{Ruan:2009ug} was proposed at the Solenoidal Tracker at RHIC (STAR) \cite{Ackermann:2002ad} to trigger on and identify muons, from which quarkonia can be reconstructed, based on its precise timing information. The di-muon channel can be used to improve precision of \jpsi\ measurement at low transverse momentum (\pT) thanks to the low kinematic threshold of the MTD trigger on muons. It also has the potential to separate different \ups\ states for the first time at STAR as the Bremsstrahlung radiation for muons is much smaller compared to electrons. In this paper, we will present the measurements of \jpsi\ suppression, \vtwo, as well as \ups\ reconstruction in Au+Au collisions at \sqrtsNN\ = 200 GeV.

\section{Analysis setup}
\label{sect::setup}
Data used in this analysis were recorded in 2014 at RHIC for Au+Au collisions at \sqrtsNN\ = 200 GeV after the MTD was fully installed earlier that year. A MTD dimuon trigger, which requires at least two hits in the MTD, was designed to trigger on quarkonium events. It sampled an integrated luminosity of 14.2 nb$^{-1}$, but only 30\% of the statistics was used for the results presented here except for the \vtwo\ measurement, which utilizes the di-electron channel using data from 2010 and 2011.

Muon candidates are identified using the Time Projection Chamber (TPC) and MTD. The TPC is the main tracking device to measure the momentum and ionization energy loss ($dE/dx$) of charged particles, with coverage of $0<\varphi<2\pi$ and $|\eta|<1.0$. Charged tracks are required to have \pT\ above 1.5 \gev, and lose energy within [-1.5$\sigma$,2.5$\sigma$] difference to the expected energy loss for muons, where $\sigma$ is the $dE/dx$ resolution. Tracks also need to geometrically match to the hits measured by the MTD, which covers about 45\% in azimuth within $|\eta|<0.5$. \pT\ dependent cuts are applied to the residuals between projected track positions and MTD hit positions along both $z$ and $\varphi$ directions. In addition, the difference between the expected time-of-flight for tracks from primary vertex to the MTD and the time measured by the MTD should not exceed 1 ns.

\begin{figure}[htbp]
\centering
\begin{minipage}{.45\linewidth}
  \includegraphics[width=\linewidth]{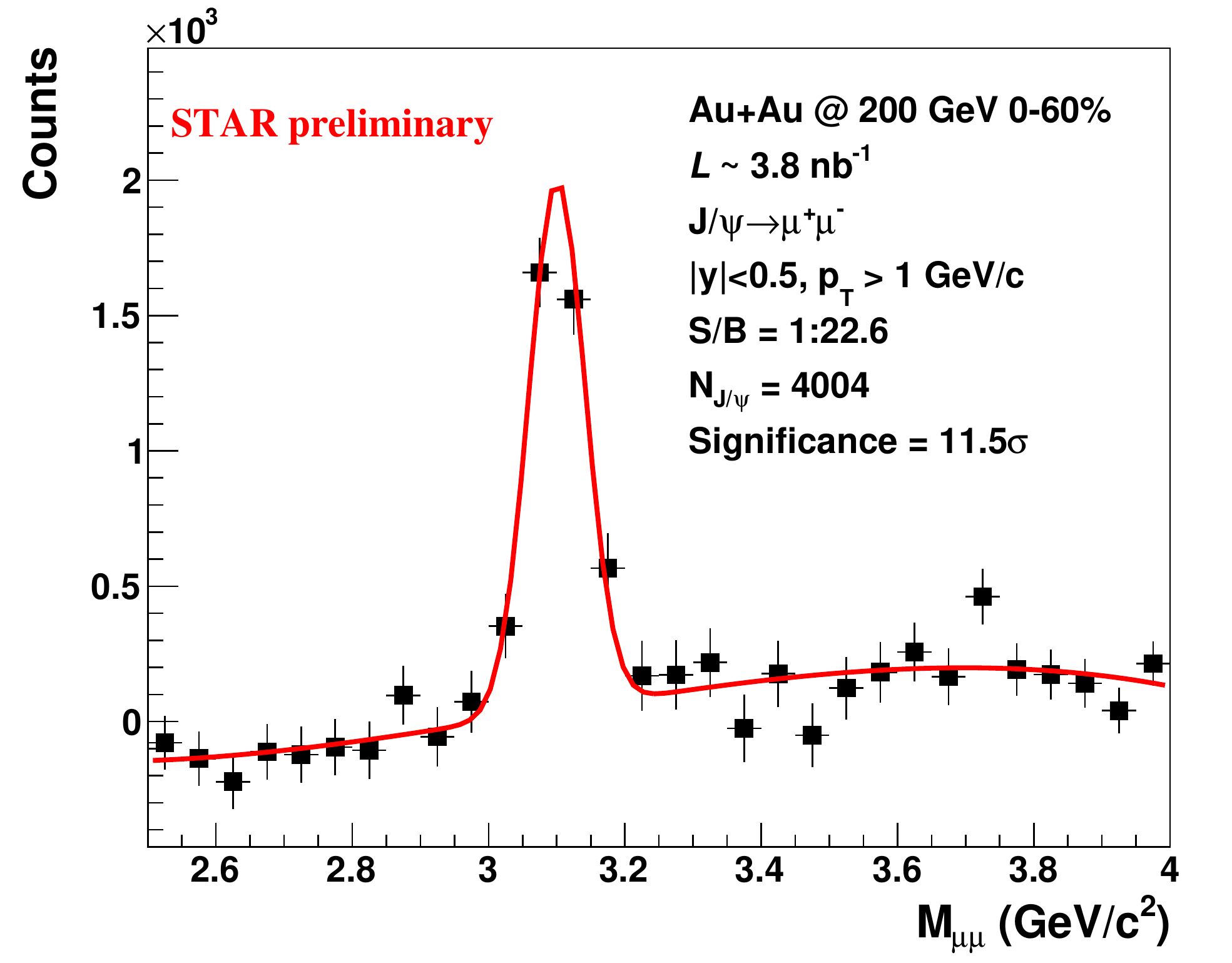}
  \caption{Invariant mass distribution of unlike-sign muon pairs after subtracting mixed-event background. A Gaussian+Pol3 fit to the distribution is shown as the red line.}
  \label{fig::signal}
\end{minipage}
\hspace{.05\linewidth}
\begin{minipage}{.45\linewidth}
  \includegraphics[width=\linewidth]{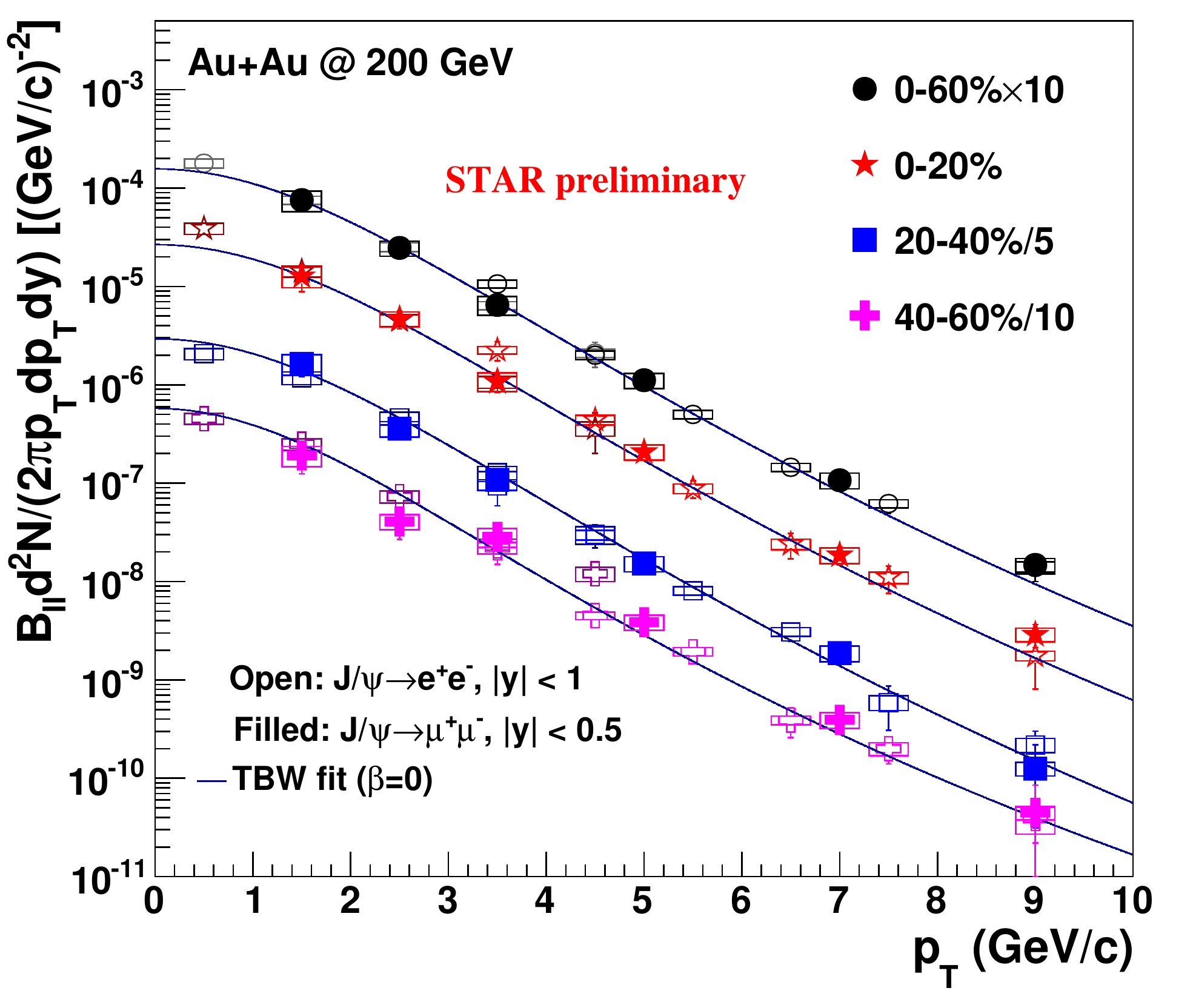}
  \caption{Cross-sections of inclusive \jpsi\ vs. \pT\ in four centralities classes (filled), with comparison to the published di-electron results (open) \cite{Adamczyk:2013tvk,Adamczyk:2012ey}. The solid lines are Tsallis Blast-wave fits to the di-electron results.}
  \label{fig::xsec}
\end{minipage}
\end{figure}

The invariant mass distribution of unlike-sign muon pairs is shown in Fig. \ref{fig::signal} as the black points after subtracting the combinatorial background estimated via event mixing, namely pairing a muon with uncorrelated muons from other events. A clear \jpsi\ peak can be seen with a total of about 4k \jpsi\ candidates.
The distribution is then fitted with a Gaussian distribution plus a third-order polynomial function shown as the red line in Fig. \ref{fig::signal}. The raw \jpsi\ yield is extracted by summing the bin contents of the invariant mass distribution within the mass range of 2.9-3.3 \gevtwo\ and subtracting the residual background estimated from the combined fit. 

\section{Results}
\label{sect::result}
After correcting for the detector inefficiency and acceptance, cross-sections of inclusive \jpsi\ as a function of \pT\ are shown in Fig. \ref{fig::xsec} as filled symbols for four centrality classes. The statistical errors are indicated as vertical bars while the systematic uncertainties are shown as open boxes. The centrality class is determined based on event multiplicity with boundaries extracted using the Glauber model \cite{Miller:2007ri} to reflect the collision geometry. Published results for low and high \pT\ \jpsi\ cross-sections via the di-electron channel \cite{Adamczyk:2013tvk,Adamczyk:2012ey} are shown as the open symbols, and agree quite well with the new di-muon results. 

The suppression of \jpsi\ yield is quantified by the nuclear modification factor (\RAA), whose definition is
\begin{equation}
\RAA = \frac{Y(\rm{AuAu})}{N_{\rm{coll}}Y(\rm{pp})}
\end{equation}
where $Y(\rm{AuAu})$ and $Y(\rm{pp})$ are the invariant yields in \AuAu\ and \pp\ collisions respectively, and \ncoll\ is the number of binary nucleon-nucleon collisions in each heavy-ion collision centrality class estimated using the Glauber model.
\begin{figure}[htbp]
\centering
\begin{minipage}{.45\linewidth}
  \includegraphics[width=\linewidth]{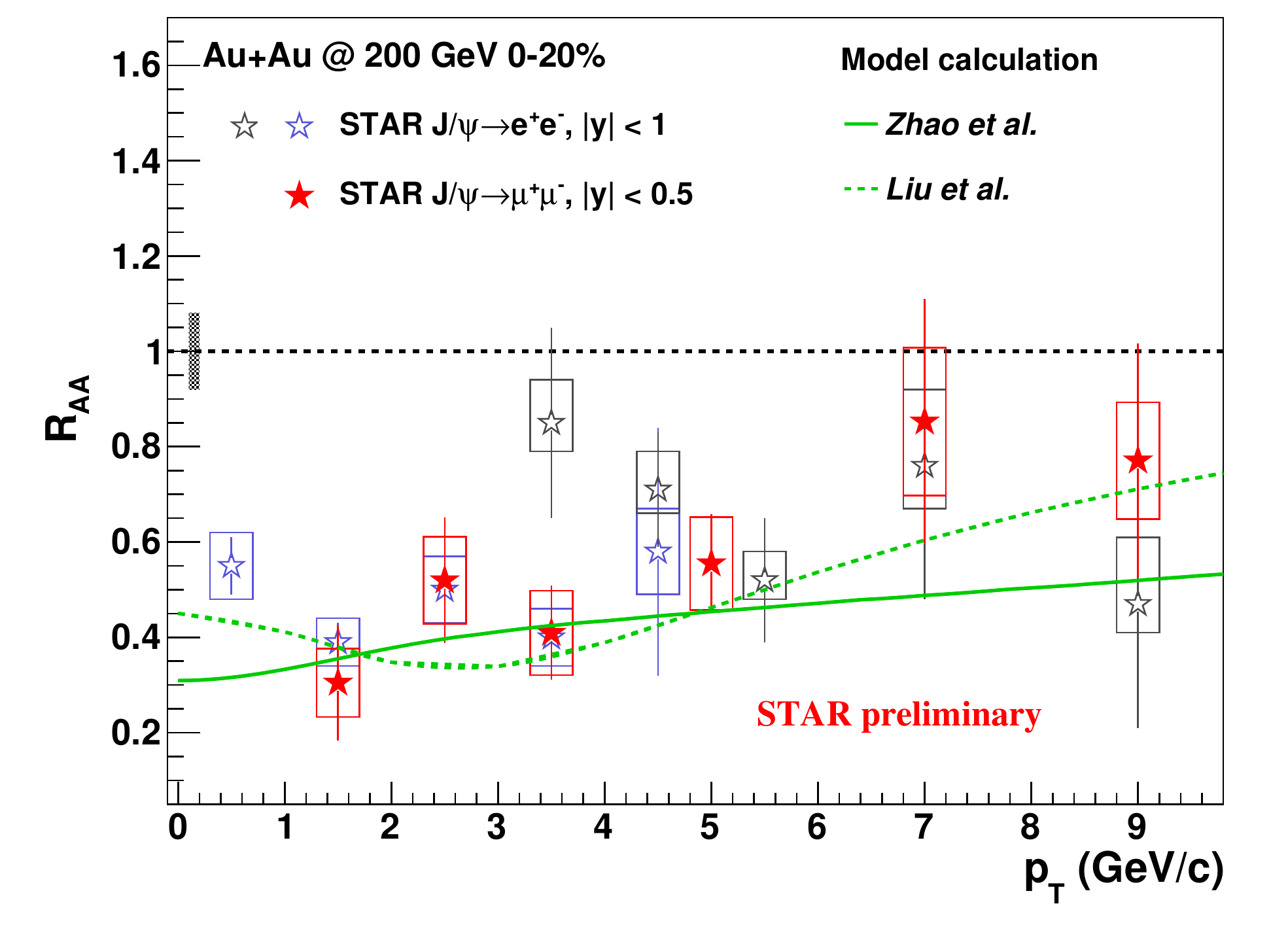}
  \caption{\jpsi\ \RAA\ as a function of \pT\ via di-muon (filled) and di-electron (open) channels \cite{Adamczyk:2013tvk,Adamczyk:2012ey} in 0-20\% most central events. Model predictions are shown as green curves \cite{Adamczyk:2012ey}. }
  \label{fig::raavspt}
\end{minipage}
\hspace{.05\linewidth}
\begin{minipage}{.45\linewidth}
  \includegraphics[width=\linewidth]{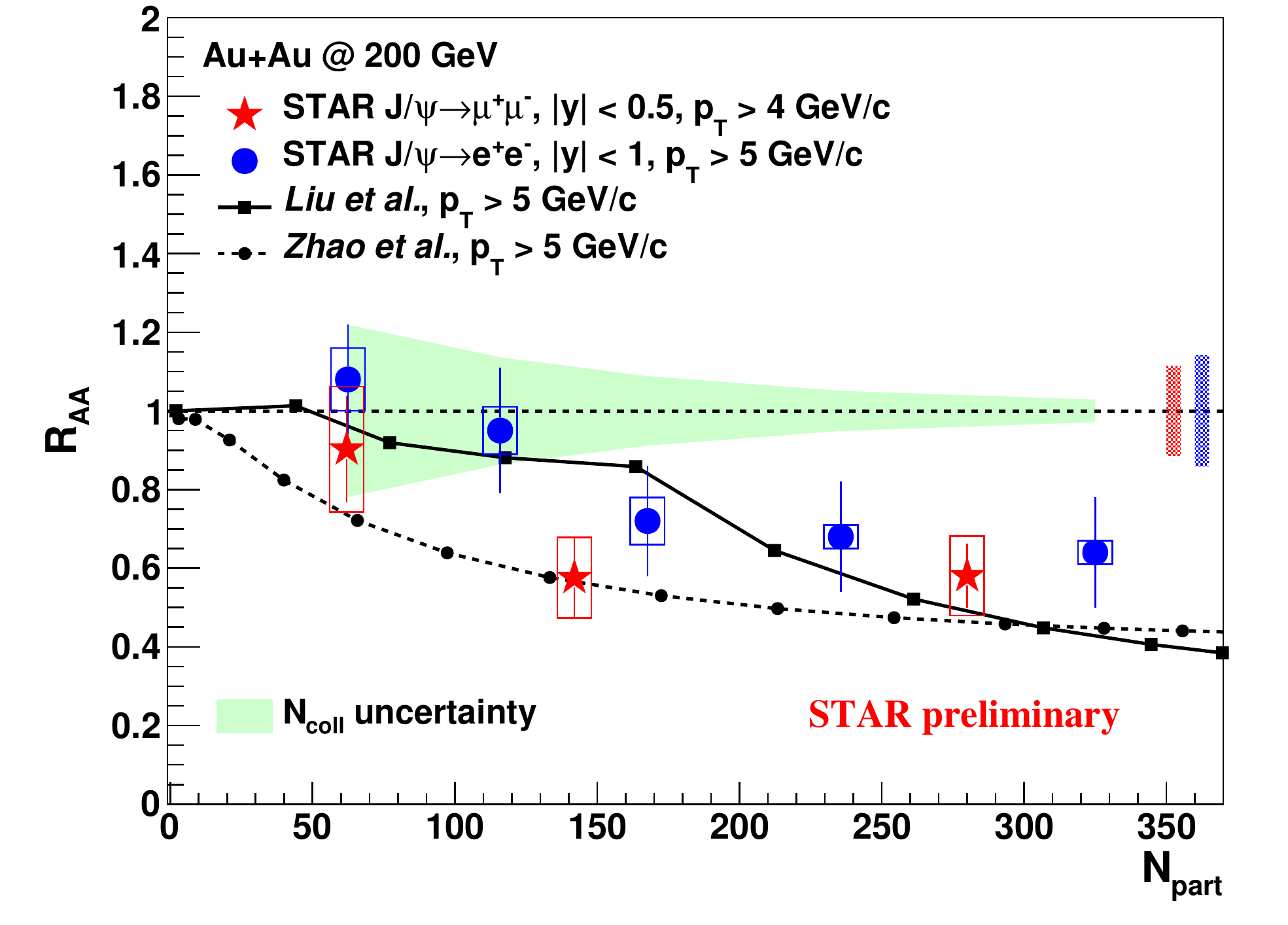}
  \caption{\RAA\ of high \pT\ \jpsi\ as a function of \npart\ for di-muon (filled) and di-electron (open) channels \cite{Adamczyk:2013tvk,Adamczyk:2012ey}. Model predictions are shown as black curves \cite{Adamczyk:2012ey}.}
  \label{fig::raavsnpart}
\end{minipage}
\end{figure}
The \jpsi\ \RAA\ is shown in Fig. \ref{fig::raavspt} as a function of \pT\ for the di-muon (filled stars) and published di-electron (open stars) channels. A distinct rising trend from low to high \pT\ is seen. The effects of dissociation, cold nuclear matter (CNM) and recombination all play a role at low \pT. On the other hand, contributions of recombination and CNM effect become negligible at high \pT. Due to the longer formation time, high \pT\ \jpsi\ is more likely to be formed outside of the medium, making it less probable to be dissociated in the medium. Furthermore, a significant fraction of inclusive \jpsi\ in this kinematic region originates from B-hadron decays. At the LHC, non-prompt \jpsi\ from B-hadron decays has been shown to be less suppressed compared to prompt \jpsi\ for $6.5<\pT<30$ \gev\ at mid-rapidity in 0-20\% central Pb+Pb collisions at \sqrtsNN\ = 2.76 TeV \cite{Chatrchyan:2012np}. All these effects can contribute to the rising trend of \RAA. However, the fact that there is still sizable suppression at high \pT\  despite the discussed effects suggests that color-screening is probably the main mechanism responsible as initially proposed. The integrated \RAA\ above 4 \gev\ is presented in Fig. \ref{fig::raavsnpart} as a function of number of participating nucleons (\npart), calculated using the Glauber model. It is $\sim0.6$ in mid-central and central collisions, consistent with the expectation of a deconfined medium created in these collisions whose temperature is high enough to dissociate \jpsi. Two model predictions incorporating both dissociation of prompt \jpsi\ and contribution of regenerated \jpsi\ can describe the measured \pT\ and \npart\ dependence of \RAA\ reasonably well (see references in \cite{Adamczyk:2012ey} for models).

\begin{figure}[htbp]
\centering
\begin{minipage}{.45\linewidth}
  \includegraphics[width=\linewidth]{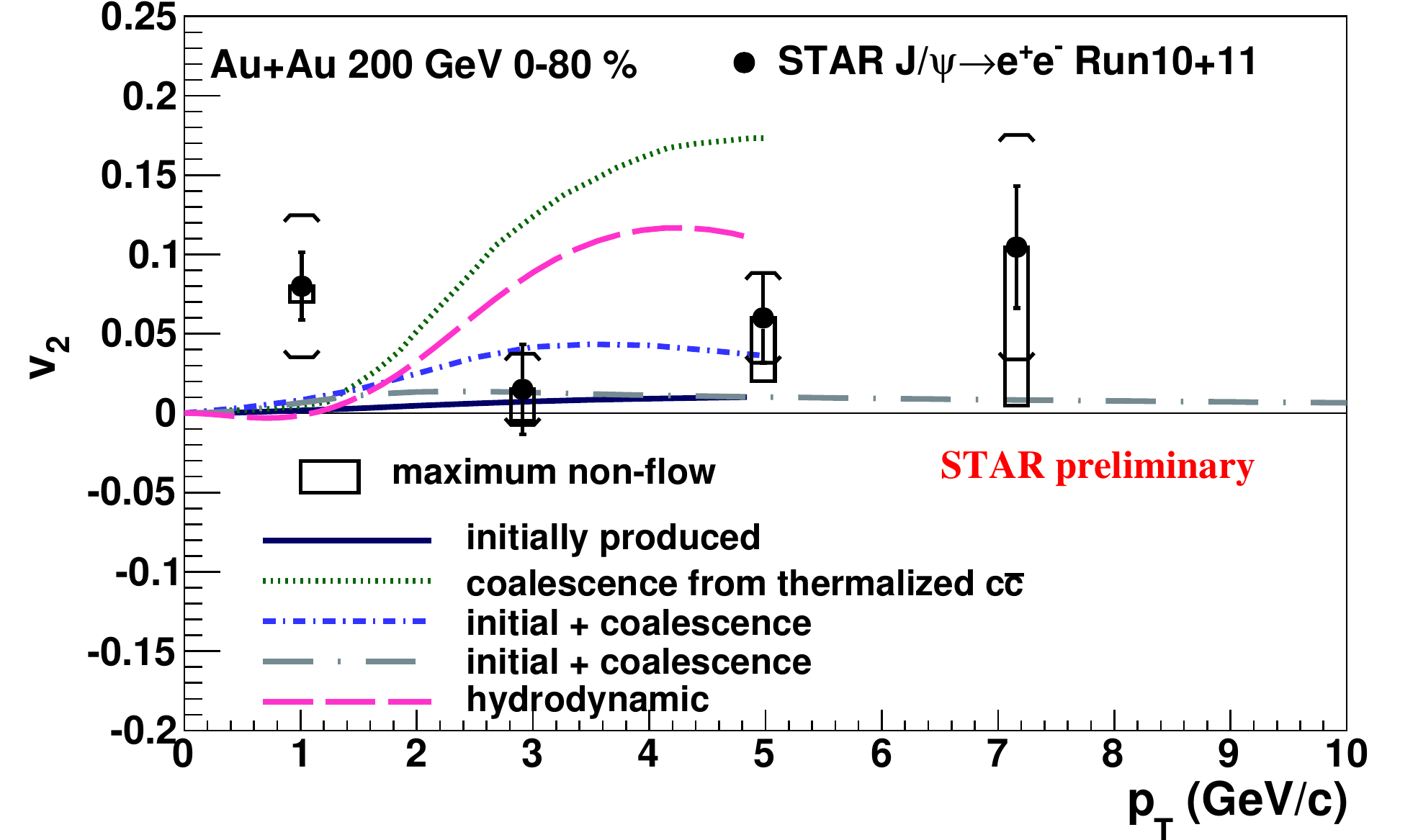}
  \caption{\jpsi\ \vtwo\ as a function of \pT\ via di-electron channel in 0-80\% \AuAu\ events. Published results \cite{Adamczyk:2012pw} using data taken in 2010 are combined with the new results using 2011 data. }
  \label{fig::v2vspt}
\end{minipage}
\hspace{.05\linewidth}
\begin{minipage}{.45\linewidth}
  \includegraphics[width=\linewidth]{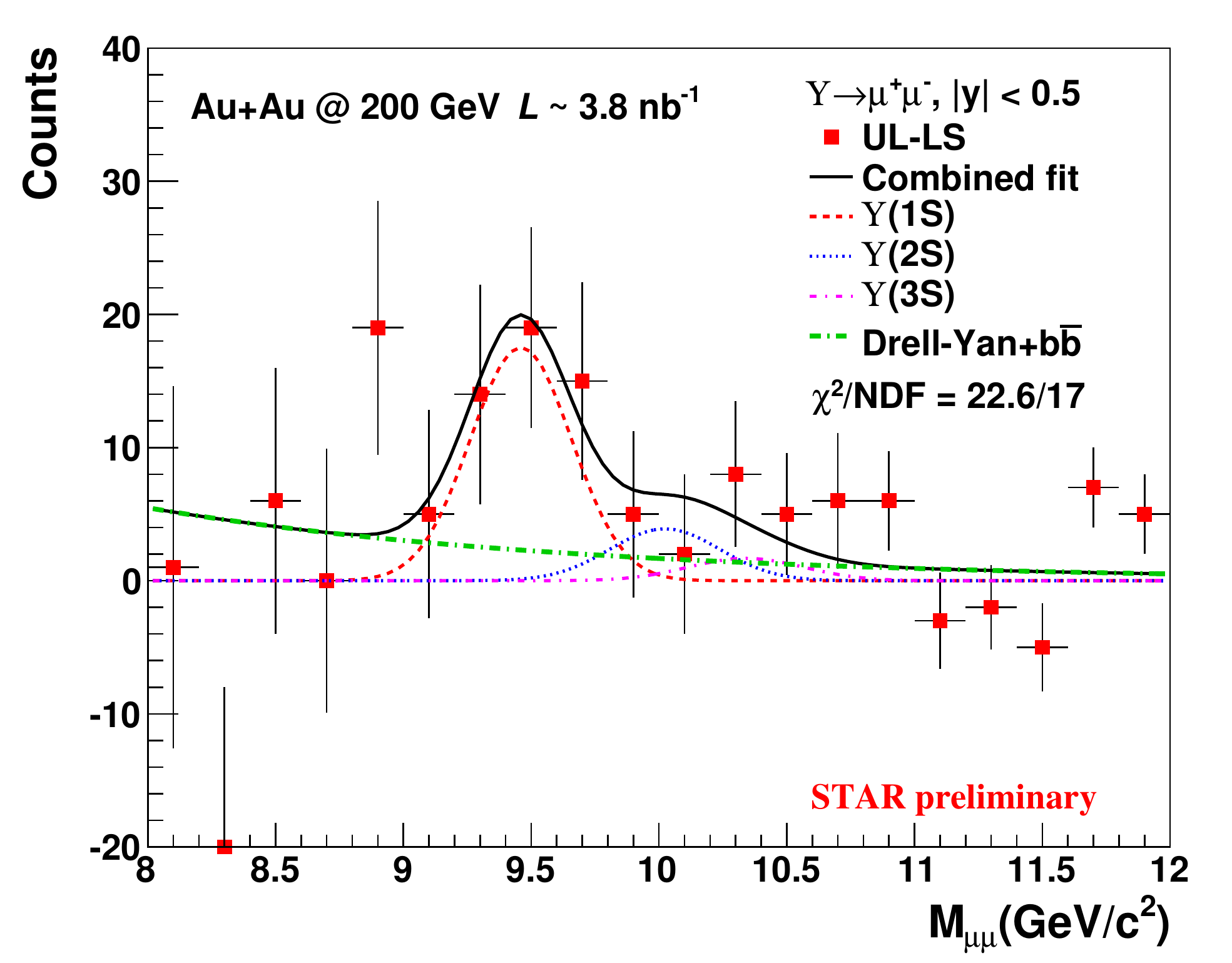}
  \caption{Invariant mass distribution of unlike-sign dimuon pairs after subtracting the like-sign distribution (red points). A combined fit of signal and residual background is shown as the solid black curve.}
  \label{fig::upsilon}
\end{minipage}
\end{figure}

To further discriminate different production mechanisms, \jpsi\ \vtwo\ is measured via the di-electron channel as shown in Fig. \ref{fig::v2vspt} where the published results using  2010 data \cite{Adamczyk:2012pw} are combined with new results using 2011 data. The data are shown as black points along with the statistical errors (vertical bars) and systematic uncertainties (brackets), while the non-flow contribution is shown as the open boxes. For \pT\ above 2 \gev, the measured \vtwo\ is consistent with zero, disfavoring the scenario that the dominant contribution to the observed yields at this kinematic region is the regeneration of fully thermalized charm quarks. Models consisting of both primordial and regenerated \jpsi\ can qualitatively reproduce data (see references in \cite{Adamczyk:2012pw} for models).

\ups\ mesons are also reconstructed via the di-muon channel as shown in Fig. \ref{fig::upsilon}, where the like-sign distribution representing the background is subtracted from the unlike-sign distribution. A combined fit is shown as the black curve, where the residual QCD background shape is estimated through PYTHIA \cite{Sjostrand:2006za} and the line-shapes of the \ups\ states are determined by embedding simulated signals into real data. A total of $50\pm22$ \ups's are observed, and a factor of 6 more statistics is expected combining the rest of 2014 data and the data that will be taken in 2016.

\section{Summary}
We present the first measurements of quarkonium production via the di-muon channel at mid-rapidity by the STAR experiment, thanks to the new MTD, in \AuAu\ collisions at \sqrtsNN\ = 200 GeV. The cross-sections of inclusive \jpsi\ are measured within $1<\pT<10$ \gev\ for different centrality classes. Integrated \RAA\ is  $\sim0.6$ for \pT\ above 4 \gev\ in central collisions, suggesting that the dissociation mechanism is in play. The \vtwo\ is consistent with zero for \pT\ above 2 \gev, disfavoring a dominant contribution of regeneration from fully thermalized charm quarks. The \ups\ measurement is also explored with the small data sample, and more statistics is expected.




\bibliographystyle{elsarticle-num}
\bibliography{RMa_QM2015}







\end{document}